\documentclass[fleqn,usenatbib]{mnras}

\usepackage{newtxtext,newtxmath}
\usepackage[T1]{fontenc}

\DeclareRobustCommand{\VAN}[3]{#2}
\let\VANthebibliography\thebibliography
\def\thebibliography{\DeclareRobustCommand{\VAN}[3]{##3}\VANthebibliography}

\usepackage{graphicx}	% Including figure files
\usepackage{amsmath}	% Advanced maths commands
\usepackage{orcidlink}  %ORCID LOGO
\usepackage{threeparttable} %table footnote
\title[The 2019 spin-up episode of GX 301-2]{Changes in the distribution of circum-binary material around the HMXB GX 301-2 during a rapid spin-up episode of the neutron star}

\author[Hemanth M et al.]{
Hemanth Manikantan,$^{1}$\thanks{E-mail: hemanthm@rri.res.in}${\orcidlink{0000-0001-9404-1601}}$
Biswajit Paul,$^{1}$
Kinjal Roy,$^{1}$ ${\orcidlink{0000-0002-7391-5776}}$
and Vikram Rana$^{1}$ ${\orcidlink{0000-0003-1703-8796}}$
\\
$^{1}$Raman Research Institute, C V Raman Avenue, Sadashivanagar, Bangalore-560080, Karnataka, India
}

\date{Accepted 2022 December 24. Received 2022 December 6; in original form 2022 September 15}

\pubyear{2022}

\begin{document}
\label{firstpage}
\pagerange{\pageref{firstpage}--\pageref{lastpage}}
\maketitle

\begin{abstract}
Some accretion powered X-ray pulsars with supergiant companion stars undergo occasional rapid spin-up episodes that last for weeks to a few months. We explore the changes in the accretion environment of the pulsar GX 301-2 during its latest 80 days long spin-up episode in 2019 when the spin frequency of the pulsar increased by $\sim$2\% over two orbits of the binary. By performing time-resolved spectroscopy with the \textit{MAXI}/GSC spectra of the source, we estimated the equivalent hydrogen column density and equivalent width of the iron fluorescence line during the spin-up episode, and compared them with the long-term average values estimated by orbital-phase resolved spectroscopy. The measured absorption column density during the spin-up episode is about twice that of an average orbit, while the equivalent width of the iron line is less than half of an average orbit. Though the spin-up episode started immediately after a pre-periastron flare and lasted for the two consecutive orbits of the binary, the associated enhancement in luminosity started a few days after the pre-periastron flare and lasted only during the first orbit, and some enhancement was seen again during the pre-periastron passage of the second orbit. The absorption column density and iron line equivalent width vary throughout the spin-up episode and are distinct from an average orbit. These observations indicate a significant change in the accretion and reprocessing environment in GX 301-2 during the spin-up episode and may hold important clues for the phenomenon in this source and several other sources with supergiant companions.
\end{abstract}

\begin{keywords}
accretion, accretion discs -- stars: neutron -- pulsars: general -- X-rays: binaries -- X-rays: individual: GX 301-2
\end{keywords}

%%%%%%%%%%%%%%%%%%%%%%%%%%%%%%%%%%%%%%%%%%%%%%%%%%

%%%%%%%%%%%%%%%%% BODY OF PAPER %%%%%%%%%%%%%%%%%

\section{Introduction} \label{intro}

GX 301-2 (4U 1223-62) is an accreting X-ray pulsar in an eccentric ($e\sim0.472$) binary orbit around the 62 R$_\odot$ and $\ge35$ M$\odot$ early type supergiant Wray 15-977 (spectral type B1.5 Ia+) (\citealt{koh_gx301_spinup, kaper2006vlt}). A peculiar characteristic of the pulsar in this HMXB is that it exhibits a significant increase in X-ray intensity $\sim1.4$ days before the periastron passage of the pulsar, periodically in 41.5 days intervals which is the orbital period of the binary. The inclination angle is estimated to be in the range of 68-78$^{\circ}$ based on the absence of eclipses and estimates of the size and mass of the companion \citep[See][and references therein]{sato1986orbital}. The presence of characteristic emission lines of iron at $6.4$ keV (Fe K$\alpha$) and its Compton shoulder in the observed spectrum indicate the presence of dense stellar wind from the companion, which reprocesses the X-rays from the pulsar \citep[See][and references therein]{watanabe2003detection}. Since a spherically symmetric stellar wind from the companion would only endorse a periastron flare, several models have been proposed to explain the unusual pre-periastron flaring nature of the source \citep{pravdo_2001ApJ...554..383P,leahy_2008MNRAS.384..747L}. In addition to the strong, symmetric, high-velocity stellar wind from the massive companion star, the model by \cite{leahy_2008MNRAS.384..747L} proposed a dense equatorial gas stream from Wray 15-977. On the other hand, \cite{pravdo_2001ApJ...554..383P} suggested the presence of an enhanced circumstellar disc around Wray 15-977. Both models predict the orbital variation of the observed X-ray flux and the equivalent hydrogen column density. Along with the observed column density, the equivalent width of the iron fluorescence line at different orbital phases is another diagnostic tool for probing the accretion/reprocessing environment of the neutron star in the circum-binary material. The orbital profile of the intensity, absorption column density and equivalent width were measured with the first few years of data obtained with \textit{MAXI}/GSC \citep{nazma2014} and the results, however, were found to deviate from the predictions made by these two models.

\begin{figure}
	\includegraphics[width=\columnwidth]{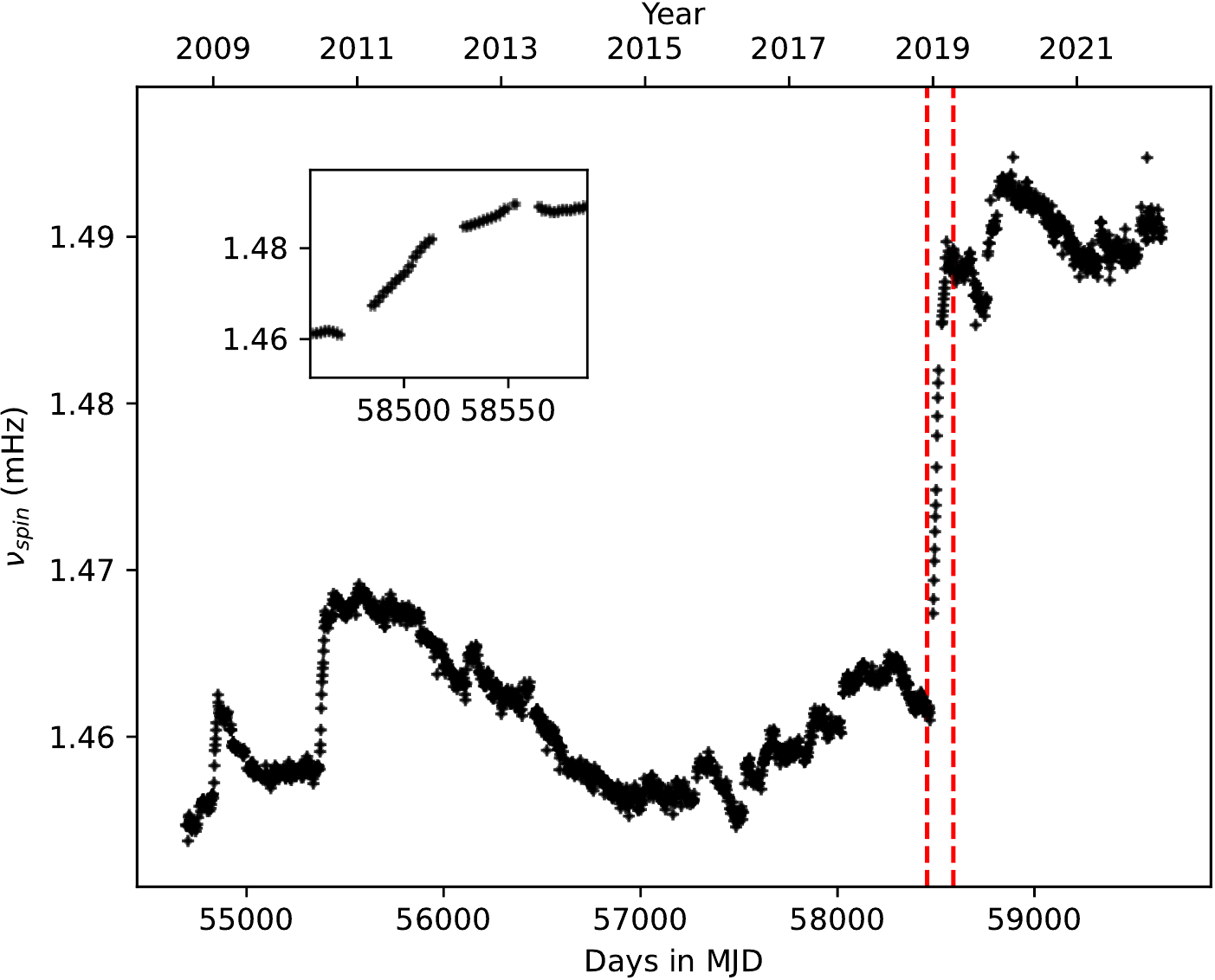}
    \caption{Figure showing the pulse frequency history of GX 301-2 obtained with \textit{Fermi}/GBM. The red dashed line indicates the 2019 spin-up episode of GX 301-2. The spin-up episode roughly spans 80 days between MJD 58480 to MJD 58560 and is shown in the inset.}
    \label{fig:spin-up_episode}
\end{figure}

GX 301-2 is peculiar in its spinning nature as well, that it exhibits rapid spin-up episodes, which is an observed characteristic of accreting X-ray pulsars (XRPs) in Be high mass X-ray binaries (Be HMXBs), but it also exhibits transient spin-ups and spin-downs, which is a characteristic of XRPs in Sg HMXBs \citep[See][]{bildsten1997observations}. During a rapid spin-up episode in 2019, the spin frequency of the pulsar in GX 301-2 increased by about 2\% in about 80 days (See \citealt{nabizadeh2019}, \citealt{abarr2020observations}, \citealt{liu2020spin}, \citealt{liu2021disc} and Figure. \ref{fig:spin-up_episode} of this paper), accompanied by an overall increase in the X-ray luminosity. The pulsar exhibited an enhanced X-ray flux in both low (2-30 keV) as well as high (15-50 keV) energies evident from \textit{MAXI} and \textit{Swift}/BAT light-curves respectively. While \citealt{nabizadeh2019} and \citealt{liu2021disc} used the pointed observations of \textit{NuSTAR} and \textit{Insight-HXMT} respectively during the 2019 spin-up episode and outside the spin-up episode to probe the differences in the spectral properties, \citealt{abarr2020observations} reported constraints on the hard X-ray polarization using \textit{X-Calibur}, and \citealt{liu2020spin} used the \textit{Fermi}/GBM and \textit{Swift}/BAT data to suggest the presence of a transient accretion disk during the spin-up.

In this study, we used the long-term data of the source acquired by \textit{MAXI}/GSC. We performed an orbital phase resolved spectroscopy of the source to improve the measurements of the X-ray reprocessing environment of the source using 14 years of \textit{MAXI}/GSC data from 2008 to 2022. We further performed a time resolved spectroscopy of the source during the spin-up episode to investigate changes in the accretion/reprocessing environment of the neutron star during the long spin-up episode in 2019.

\section{Methods, Instruments, Observations, Spectral Analysis}

\subsection{Methods}
Probing the strength of the photoelectric absorption (the absorption column density) and the iron fluorescence line (the equivalent width) in the spectra is a useful tool to understand the X-ray reprocessing environment of the pulsar \citep[See][and references therein]{nazma2014}. Both of these parameters can be estimated by analysing the X-ray spectra of the source.

In order to estimate the orbital variation of spectral parameters, it is desirable to have pointed observations of the source covering an entire orbit of the source. However, the binary orbital period of GX 301-2 is long at about 41.5 days \citep{sato1986orbital}, and pointed observations at the source spanning complete binary orbit(s) are not feasible. \textit{MAXI} on the other hand being an All-sky monitor scans the entire sky every day, and its spectral coverage at low energies helps constrain the photoelectric absorption parameter. The orbital variation of spectral parameters of GX 301-2 using $\sim4$ years of data accumulated by \textit{MAXI}/GSC was reported earlier in \citeauthor{nazma2014} in 2014. As of now, \textit{MAXI} have acquired almost two times more data ($\sim12$ years since MJD 55058). Hence, one should be able to constrain the orbital phase resolved spectral parameters to a greater degree by using the three times longer exposure.

Like the similar wind accreting pulsar OAO 1657-415 \citep{Jenke_2012}, GX 301-2 has also exhibited several rapid spin-up episodes throughout the time it was monitored by the all-sky monitors \textit{CGRO}/BATSE and \textit{Fermi}/GBM. Burst and Transient Source Experiment (BATSE) instrument onboard the Compton Gamma Ray Observatory (CGRO) monitored two such episodes between 1991 to 1995 \citep{koh_gx301_spinup}. \textit{Fermi}/GBM between 2008 to 2022 has detected four such episodes \citep{liu2020spin} as shown in the Figure.\ref{fig:spin-up_episode}. The longest spin-up episode of the source was the one in 2019 (vertical dashed lines in Figure.\ref{fig:spin-up_episode}), during which there were no pointed observations of the source with any of the current X-ray telescopes. We have therefore used data from the \textit{MAXI}/GSC to probe the evolution of spectral parameters over time considering the bright nature of the source. The spectrum accumulated in one day though, has very limited photon statistics to perform such an analysis. We, therefore, used the \textit{MAXI}/GSC spectra accumulated from 6-day sliding windows during the spin-up episode, to search for the temporal variation of the spectral parameters of interest.

All the spectral analysis were performed in {\small XSPEC} v12.11.1 \citep{xspec}. Interstellar photoelectric absorption is modelled using \texttt{tbabs}, with elemental abundances from \cite{wilm_2000ApJ...542..914W} and photoelectric absorption cross-sections from \cite{vern1996ApJ...465..487V}.

\subsection{Instrument and Data Reduction}

The Monitor of All-sky X-ray Image (\textit{MAXI}) \footnote{\url{https://iss.jaxa.jp/en/kiboexp/ef/maxi/}} is an X-ray scanning sky monitor installed on the Japanese Experiment Module Exposed Facility (Kibo-EF) onboard the International Space Station (ISS). \textit{MAXI} \citep{10.1093/pasj/61.5.999} is equipped with two different kinds of X-ray detectors; A pair of Gas Slit Camera (GSC) \citep{mihara2011gas} which are position-sensitive gas proportional counters detecting events in 2-30 keV and Solid-state Slit Camera (SSC) \citep{tomida2011solid}, the CCDs detecting events in 0.5-12 keV. The GSC detectors are paired to two Slit-camera optics scanning the sky in the Earth-horizon and zenith directions with the ISS motion. The 96-minute orbit of ISS around the earth facilitates GSC and SSC to scan the entire sky during one orbit. The motion of X-ray sources on the orthogonal cameras due to the ISS orbit enables the determination of source location in the sky.

The \textit{MAXI}/GSC consists of 12 position-sensitive proportional counter detectors with a total geometric area of about 5300 cm$^2$, has a FOV of $1.5^{\circ}\times160^{\circ}$  and provides an energy resolution of 18\% FWHM at 5.9 keV. In this work, we have used the GSC data for performing the spectral analysis owing to its superior effective area over SSC. The spectra and response files were downloaded from MAXI on-demand process\footnote{\url{http://maxi.riken.jp/mxondem/index.html}}.

The Gamma-ray Burst Monitor (GBM) is the secondary instrument onboard the Fermi Gamma-ray Space Telescope \citep{fermi}. GBM comprises 12 Sodium Iodide (NaI) scintillation detectors and 2 Bismuth Germanate (BGO) Scintillation detectors covering the energy ranges of 8 keV to 1 MeV and 150 keV to 40 MeV respectively. The data acquired by the GBM NaI detectors in CTIME data mode with 0.256 s time resolution in the 12-50 keV energy band is used  by the GBM Accreting Pulsars Program to measure the spin frequency of accreting X-ray pulsars. Typically, the integration times ranging from 1 to 4 days are searched for pulse frequency using the epoch folding technique and corrected for the solar system barycenter and for the known binary orbit of the source. A review of the GBM Accreting Pulsars Program is present in \cite{Malacaria_2020}. The spin history of GX 301-2 was downloaded from GBM Accreting Pulsar Histories\footnote{\url{https://gammaray.nsstc.nasa.gov/gbm/science/pulsars.html}}.

The Burst Alert Telescope (BAT) on the Neil Gehrels Swift Observatory \citep{swiftbat} is a hard X-ray All-sky monitor operating in the 15-150 keV energy band. \textit{Swift}/BAT consists of a combination of an array of Cadmium Zinc Telluride detectors (CdZnTe) with a total detection area of 5200 cm$^2$ and a Coded Aperture mask made of Lead tiles to image the X-ray sky in its 1.4 str field of view. Since the \textit{Swift}/BAT provides almost 88\% coverage of the sky every day \citep{bat_webKrimm_2013}, long-term light-curves of many X-ray sources have been made which are publicly available\footnote{\url{https://swift.gsfc.nasa.gov/results/transients/}}. We have used the \textit{Swift}/BAT long-term light-curve of GX 301-2 in this work.

\subsection{Long-term averaged spectrum and average spectrum during the spin-up episode}
We fitted the 3.5-20 keV \textit{MAXI}/GSC spectrum of GX 301-2 with different models, all of them including an iron line and: i) a power-law, ii) power-law with a high energy cutoff, and iii) power-law with a partial covering absorption and iv) power-law with a high energy cutoff, with a partial covering absorption.

\begin{figure}
    \centering
    \includegraphics[width=\linewidth]{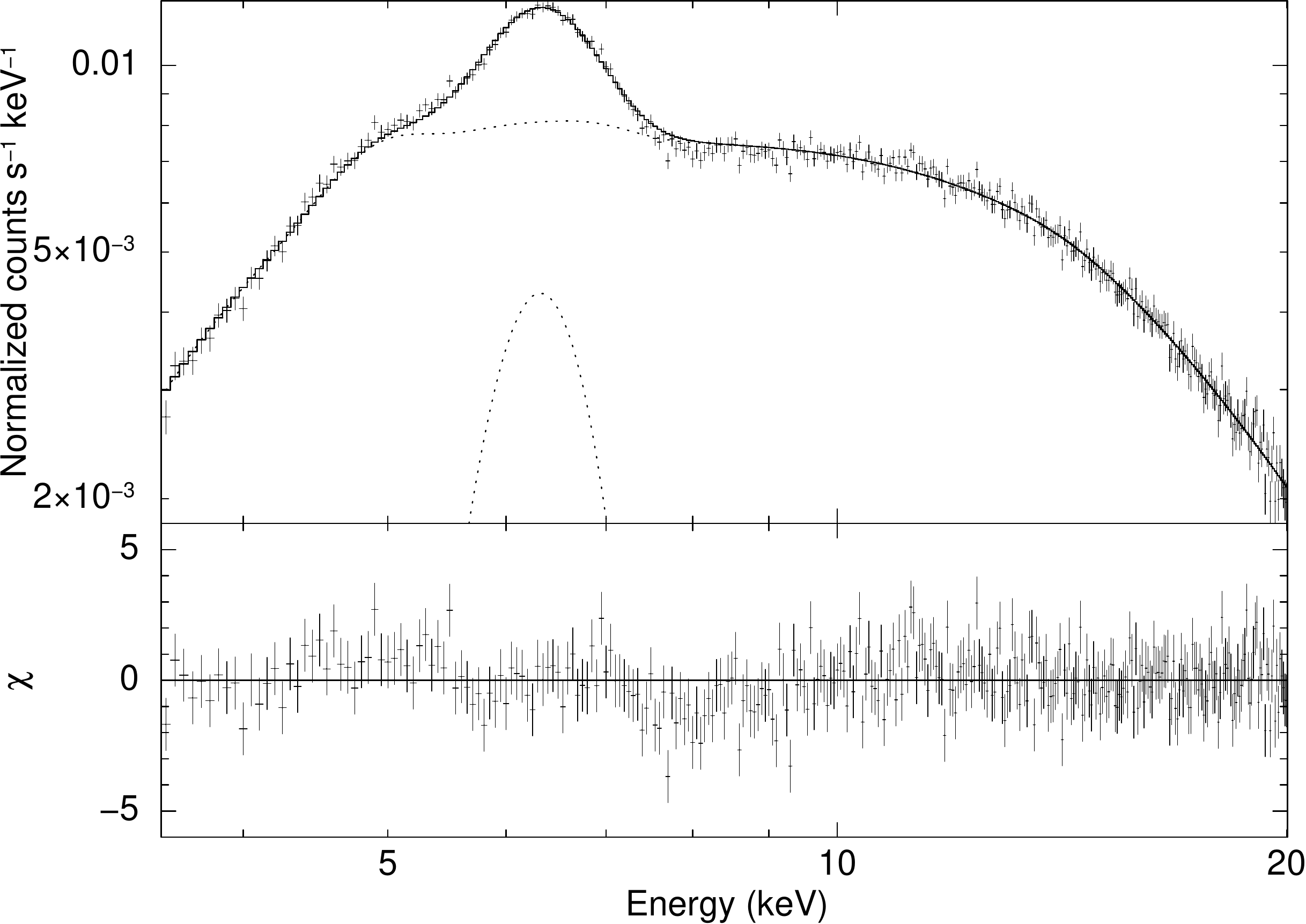}
    \includegraphics[width=\linewidth]{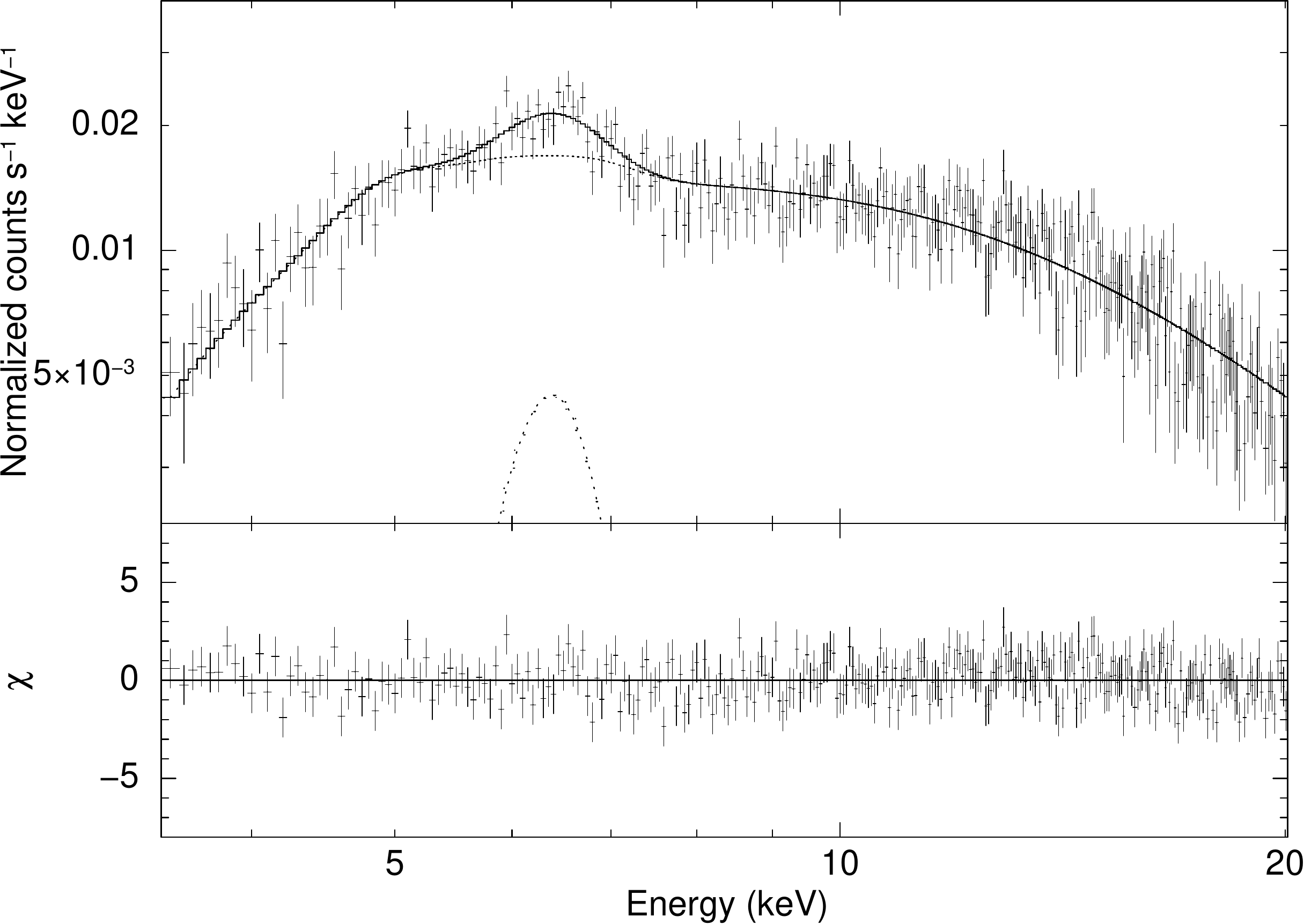}
    \caption{The figure on the top shows the \textit{MAXI}/GSC spectrum and the best fit spectral model \texttt{tabs(powerla$*$highecut+gauss)} for the long term average spectrum of GX 301-2. The figure on the bottom shows the \textit{MAXI}/GSC spectrum and the best fit model \texttt{tabs(powerla+gauss)} during the spin-up episode in MJD 58480-58560. In both the figures, the top panel shows the spectrum and the best fit model, while the bottom panel shows the residuals to the best fit model.}
    \label{fig:orb_avg_spectrum}
\end{figure}

Though short observations of GX 301-2 are known to exhibit partially covered power-law with cutoff  at high energies, with fluorescent emission lines of iron \citep{umukherjee}, the long-term average \textit{MAXI}/GSC spectrum does not require a partial covering absorption. We obtained a good fit with a model consisting of an absorption column, power-law with a high energy cutoff and a gaussian emission line of iron. The best fit spectral parameters are given in Table.  \ref{tab:orb_avg_spectrum} and the spectrum along with the residuals to the best fit is shown in Figure. \ref{fig:orb_avg_spectrum}.

The long-term averaged spectrum is dominated by the relatively bright main peak of the X-ray orbital intensity profile (Near orbital phase 0.95 shown in the top panel of Figure. \ref{fig:oprs_3.5-20keV_hec_spectrum}). The iron line has an equivalent width (eqw\textsubscript{Fe}) of $708\pm1$ eV, one of the highest values among HMXB systems, and the absorption column density is $16.22\pm1.22$ in units of $10^{22}$ atoms cm\textsuperscript{-2}. The best fit spectral parameters are given in Table.  \ref{tab:orb_avg_spectrum} and shown in Figure. \ref{fig:orb_avg_spectrum}.

We fitted the overall \textit{MAXI}/GSC spectra of the source in MJD 58480-58560 during the spin-up episode and found that the model \texttt{tbabs(powerlaw+gaus)} gives a good fit. All the parameters except the gaussian width were constrained by the fit, and the best fit spectral parameters are listed in the Table.  \ref{tab:orb_avg_spectrum}. The spectra, the best fit model and residuals to the best fit model are shown in Figure. \ref{fig:orb_avg_spectrum}.

\begin{table*}
    \centering
    \caption{The table lists the best fit spectral fit parameters for the long term average spectrum and the spectrum during the 2019 sin-up episode. The errors quoted on all the spectral parameters are their 90\% confidence ranges.}
    \begin{tabular}{lllcc}
    \hline
    \hline
         Model& Parameter& Units &Long term average &Spin-up episode\\
    \hline
         tbabs& N\textsubscript{H} &$10^{22}$ atoms cm$^{-2}$ &$16.22\pm1.22$ &$30.9\pm6$\\
         powerlaw &$\Gamma$ &&$0.41\pm0.04$ &$0.94\pm0.12$\\
          &Norm. &photons keV$^{-1}$ cm$^{-2}$ s$^{-1}$ &$0.024\pm0.003$ &$0.16\pm0.05$\\
         highecut &E\textsubscript{cut} &keV &$13.95\pm0.41$ &-\\
          &E\textsubscript{fold} &keV &$15.61\pm1.8$ &-\\
         gaussian &E\textsubscript{line} &keV &$6.29\pm0.01^{\dagger}$ &$6.34\pm0.11$\\
          &$\sigma$\textsubscript{line} &keV &$0.26\pm0.03$ &$0.008\pm0.008$\\
          &Norm. &photons cm$^{-2}$ s$^{-1}$ &$0.008\pm0.0003$ &$0.008\pm0.002$\\
          &Eq. width & eV &$708\pm1$ &$295\pm77$\\
         Flux\textsubscript{2-20 keV} & &$10^{-9}$ ergs cm$^{-2}$ s$^{-1}$ &$2.22\pm0.001$ &$3.77\pm0.09$\\
         $\chi^2$/dof && &372.2/322 &308.8/303\\
    \hline
    \hline
    \end{tabular}
    \begin{tablenotes}
        \item $\dagger$ The slightly lower than expected measured value of E\textsubscript{line} is possibly due to the presence of iron K$\alpha$ Compton shoulder and the poor energy resolution of the \textit{MAXI}/GSC detectors.
    \end{tablenotes}
    \label{tab:orb_avg_spectrum}
\end{table*}

\subsection{Orbital phase resolved spectral analysis}
\begin{figure}
    \centering
    \includegraphics[width=\linewidth]{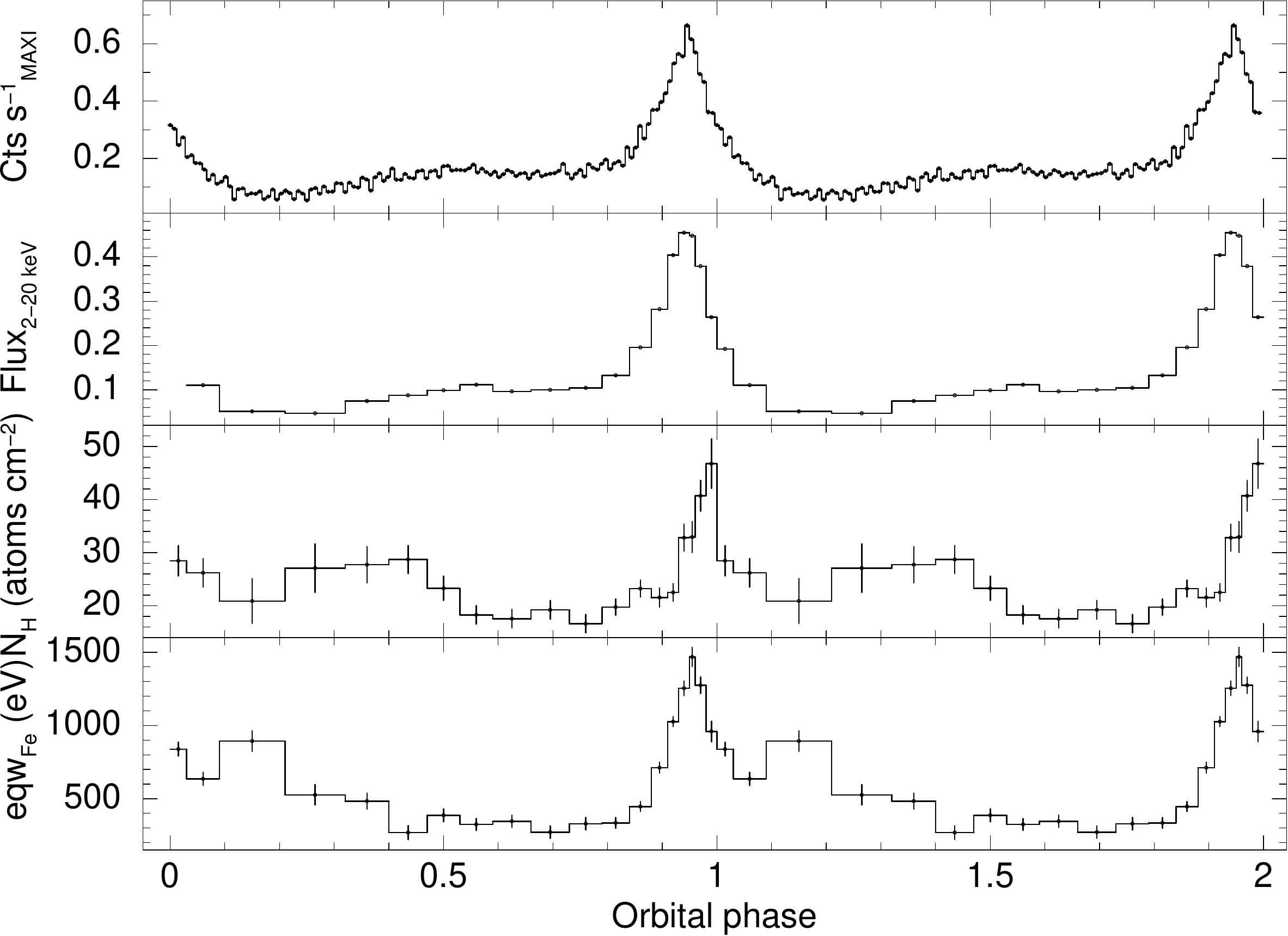}
    \caption{The top panel in the figure shows the \textit{MAXI} long-term X-ray light-curve of GX 301-2 in the 2-20 keV energy band folded with the orbital period, and the next three panels show the variation of the best fit spectral parameters over orbital phase. The 3.5-20 keV spectra in each orbital phase were fitted with the model \texttt{tbabs*(powerlaw*highecut+gaus)} and the best fit values for absorption column density and iron line equivalent width are plotted against the orbital phase. Error bars assigned to each data point are their 90\% confidence ranges. Flux\textsubscript{2-20 keV} is derived from the spectral fit and has units of photons cm\textsuperscript{-2} s\textsuperscript{-1}.}
    \label{fig:oprs_3.5-20keV_hec_spectrum}
\end{figure}

We folded the 2-20 keV \textit{MAXI} long-term light-curve with the binary orbital period (3583034 s) derived using the tool {\small efsearch} from the same light-curve at an arbitrary epoch MJD 55025.77 so that the pre-periastron peak appears at orbital phase 0.95 (aligning with \citealt{koh_gx301_spinup}). The folded light-curve (Top panel of Figure.\ref{fig:oprs_3.5-20keV_hec_spectrum}) was then split into 19 phase segments such that each phase segment has equal number of photon counts. We then generated the Good Time Intervals corresponding to these 19 orbital phase ranges within the duration of operation of \textit{MAXI}, and retrieved the spectral files corresponding to each GTI so that each of the spectra has similar photon statistics. The effective exposure durations for the spectra in each phase range are given in Table.  \ref{tab:exposure_maxi_phase}.

\begin{table}
    \centering
    \caption{The effective exposures and spectral count rates of the \textit{MAXI}/GSC spectra in each orbital phase range used in orbital phase resolved spectroscopy. }
    \begin{tabular}{lcc}
    \hline
    \hline
         Orbital phase range &Effective exposure (ks) &Count rate (Cts s\textsuperscript{-1})\\
    \hline
         0.0-0.03 &358  &$0.16\pm0.001$\\
         0.03-0.09 &723 &$0.09\pm0.001$\\
         0.09-0.21 &1271 &$0.04\pm0.001$\\
         0.21-0.32 &1216 &$0.04\pm0.001$\\
         0.32-0.40 &860 &$0.07\pm0.001$\\
         0.40-0.47 &775 &$0.07\pm0.001$\\
         0.47-0.53 &730 &$0.08\pm0.001$\\
         0.53-0.59 &703 &$0.09\pm0.001$\\
         0.59-0.66 &780 &$0.08\pm0.001$\\
         0.66-0.73 &682 &$0.08\pm0.001$\\
         0.73-0.79 &667 &$0.09\pm0.001$\\
         0.79-0.84 &609 &$0.11\pm0.001$\\
         0.84-0.88 &423 &$0.16\pm0.001$\\
         0.88-0.91 &248 &$0.24\pm0.002$\\
         0.91-0.93 &258 &$0.34\pm0.002$\\
         0.93-0.95 &179 &$0.38\pm0.002$\\
         0.95-0.96 &173 &$0.38\pm0.002$\\
         0.96-0.98 &260 &$0.32\pm0.002$\\
         0.98-1.0 &176 &$0.22\pm0.002$\\
    \hline
    \hline
    \end{tabular}
    \label{tab:exposure_maxi_phase}
\end{table}

We found that the phase resolved spectra in individual phase segments were not good enough to constrain all the spectral parameters which were well constrained in the orbital phase averaged spectroscopy. We, therefore, chose to freeze the high energy cutoff (\texttt{highecut}) parameters E\textsubscript{cut} and E\textsubscript{fold}, and the iron line parameters E\textsubscript{line} and $\sigma$\textsubscript{line} to the best fit value obtained from orbital phase averaged spectrum while performing the orbital phase resolved spectral analysis. The results of orbital phase resolved spectral analysis using power-law with a high energy cutoff are shown in the Figure. \ref{fig:oprs_3.5-20keV_hec_spectrum}. The first panel from the top shows the folded \textit{MAXI} 2-20 keV light-curve, and the second, third and fourth panels show the orbital variation of 2-20 keV flux, absorption column density and iron line equivalent width respectively, obtained from phase resolved spectroscopy.

As seen in Figure. \ref{fig:oprs_3.5-20keV_hec_spectrum}, the column density N\textsubscript{H} and eqw\textsubscript{Fe} peaks at the orbital phase corresponding to pre-periastron flare with N\textsubscript{H} reaching $\sim50\times10^{22}$ atoms cm\textsuperscript{-2} and eqw\textsubscript{Fe} $\sim1500$ eV. At around the orbital phase 0.2, eqw\textsubscript{Fe} is still high, but N\textsubscript{H} is low. Around phase 0.4, the value of N\textsubscript{H} again shows an increase to $\sim30\times10^{22}$ atoms cm\textsuperscript{-2}. These results are in agreement with \cite{nazma2014} obtained using the early \textit{MAXI}/GSC data from 2009 to 2013. However, the improved photon statistics in this study also indicate that the absorption column density is highest not during the peak of the pre-periastron flare but $1.9\pm1$ days later, during the decay phase of the pre-periastron flare.

\subsection{Time resolved spectral analysis during the spin-up episode}

Even though it is desirable to analyze the day-by-day spectrum of \textit{MAXI}/GSC to study the time-resolved spectral parameters during the spin-up episode, one-day \textit{MAXI}/GSC spectrum is limited by photon statistics. Therefore we chose multi-day sliding window spectra to fulfil the task and selected six-day windows; the window length is much less than the orbital period of the system such that orbital variation in parameters is not averaged out. Therefore we made a total of 135 sliding windows having a width of 6 days and a stride of 1 day, from MJD 58450 to 58590. This interval fully covers three orbits of the binary, including one periastron passage before the spin-up episode and two periastron passages during the spin-up episode. However, there is gap in the data especially 1) near the pre-periastron flare before the spin-up begins, and 2) just before the start of the spin-up.

\begin{figure}
    \centering
    \includegraphics[width=\columnwidth]{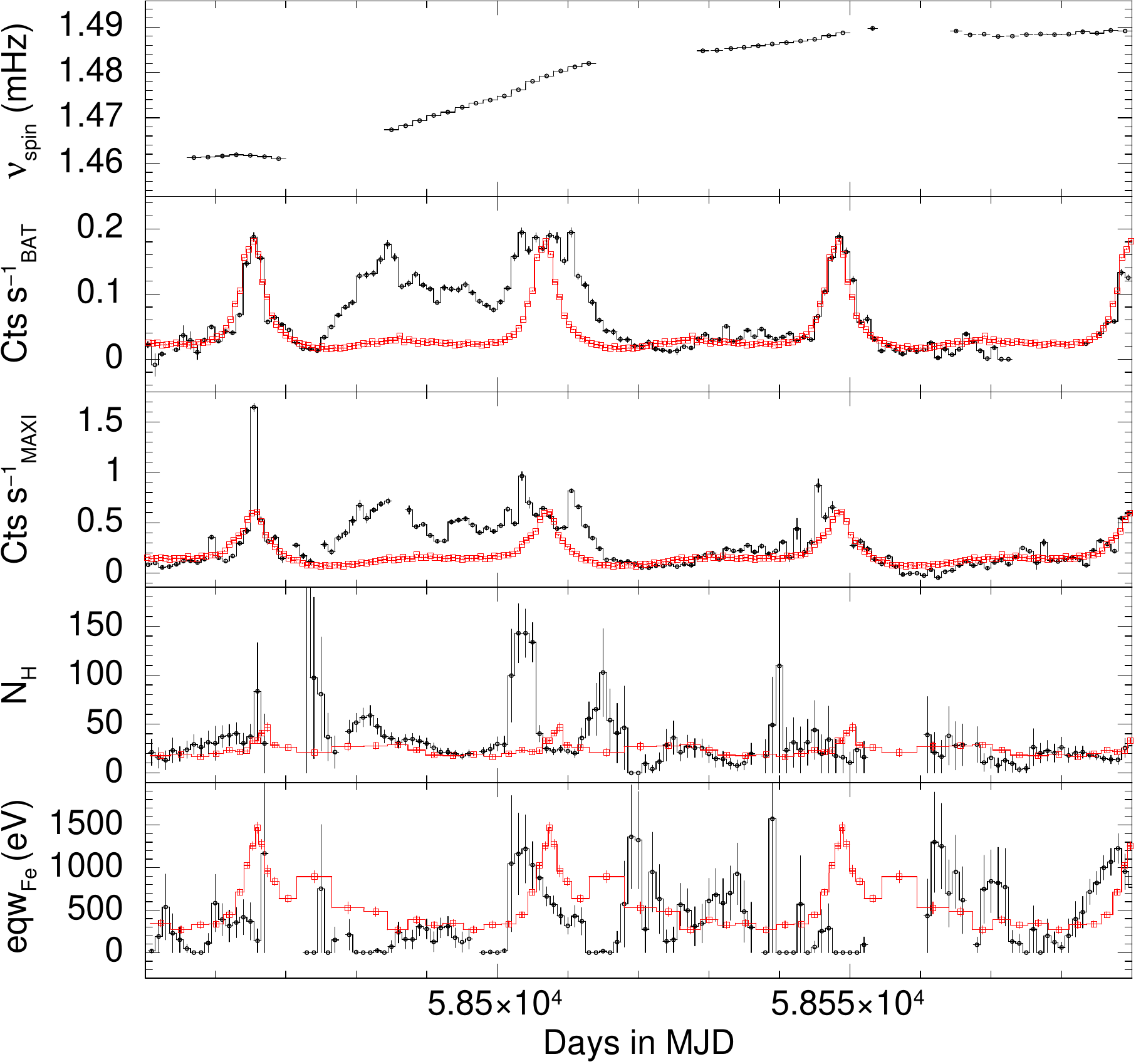}
    \caption{The data points in black shows the daily evolution of parameters while the data points in red show the long-term average orbital parameters. The first panel from the top shows the \textit{Fermi}/GBM pulsar spin history, and the next two panels show the long-term X-ray light-curves from \textit{Swift}/BAT and \textit{MAXI} and their folded profiles. Two panels from the bottom of the plot show the comparison of the spectral parameters N\textsubscript{H} (in 10\textsuperscript{22} atoms cm\textsuperscript{-2}) and eqw\textsubscript{Fe} during an average orbit to that during the spin-up episode.}
    \label{fig:compare_orb_spinup}
\end{figure}

The 3.5-20 keV spectra obtained from each of these sliding windows were then fitted with the model \texttt{tbabs(powerlaw+gaus)}. The center and width of the gaussian could not be constrained well in all the windows, and therefore center of the gaussian was fixed to the value obtained from the fit to the spectra in MJD 58480-58560, while the width of the gaussian was fixed to the value obtained from fit to the long term average spectra (See Table.  \ref{tab:orb_avg_spectrum}). The best fit values of the spectral parameters absorption column density N\textsubscript{H} and equivalent width of the iron line eqw\textsubscript{Fe} obtained from the analysis of each window are assigned to the MJD corresponding to the center of the window. The results of spectral analysis during the sliding window are shown in the Figure. \ref{fig:compare_orb_spinup}. There are three particular regions: i) high N\textsubscript{H} and low eqw\textsubscript{Fe} during MJD $\sim58480$-$58490$, ii) high N\textsubscript{H} and high eqw\textsubscript{Fe} during MJD $\sim58500$-$58510$, and iii) low N\textsubscript{H} and high eqw\textsubscript{Fe} during MJD $\sim58528$-$58538$. The \textit{MAXI}/GSC spectra in these three regions are shown in the Figure. \ref{fig:spinup_spectra_3regions}.

\section{Discussions}
The mean values and variation of the equivalent hydrogen column density and iron line equivalent width over orbital phases (Figure. \ref{fig:oprs_3.5-20keV_hec_spectrum}) are in agreement with \cite{nazma2014}, except for the secondary peak in column density near the apastron passage. We also found that the column density peaks during the decay of the pre-periastron flare and not during its maximum. Maximum absorption column density is observed $\sim1$ day after the peak of pre-periastron flare, at/near the periastron passage of the neutron star. With the improved statistics, we could also achieve considerably smaller errors on the estimates of N\textsubscript{H} and eqw\textsubscript{Fe} compared to \cite{nazma2014}. The reported discrepancy on the observations of accretion column density with the predictions from the models in \cite{pravdo_2001ApJ...554..383P} and \cite{leahy_2008MNRAS.384..747L} is very clear.

Though the \textit{Fermi}/GBM pulsar spin history has a data gap when the spin-up episode began, extrapolating the trend indicates it should have started at $\sim$ MJD 58475, about 10 days after the pre-periastron passage at $\sim$ MJD 58465. The \textit{Swift}/BAT (15-50 keV) and \textit{MAXI} (2-20 keV) light-curves also indicate an associated increase in the flux a few days after the pre-periastron flare at the same epoch (See panel 1, 2 and 3 from the top of the Figure. \ref{fig:compare_orb_spinup}). The two other spin-up episodes of GX 301-2 probed by \textit{Fermi}/GBM and \textit{Swift}/BAT also show a similar delay after the pre-periastron flare before the spin-up begins (Also see \citealt{abarr2020observations} and \citealt{liu2020spin}). The 2019 spin-up episode differs from the other spin-up episodes in the sense that it lasted for almost two binary orbits compared to less than one binary orbit for the other episodes. However, the increase in X-ray flux compared to the long-term average is seen throughout the first orbit and again during the pre-periastron flare of the second orbit. We also point out that the increase in spin-up rate happened for a luminosity comparable to the pre-periastron flare seen in every orbit of this source. GX 301-2 does not show a spin-up episode along with the increase in X-ray flux at the pre-periastron flare in every orbit. It is, therefore, likely that GX 301-2 usually has direct wind accretion, and the large spin-up episodes are instances when an accretion disk is formed around the NS. The time resolved spectral analysis presented here using the \textit{MAXI}/GSC data shows that the formation of the accretion disk is driven by some process that also causes significant changes in the X-ray reprocessing environment in the binary system.

\begin{figure}
    \centering
    \includegraphics[width=\columnwidth]{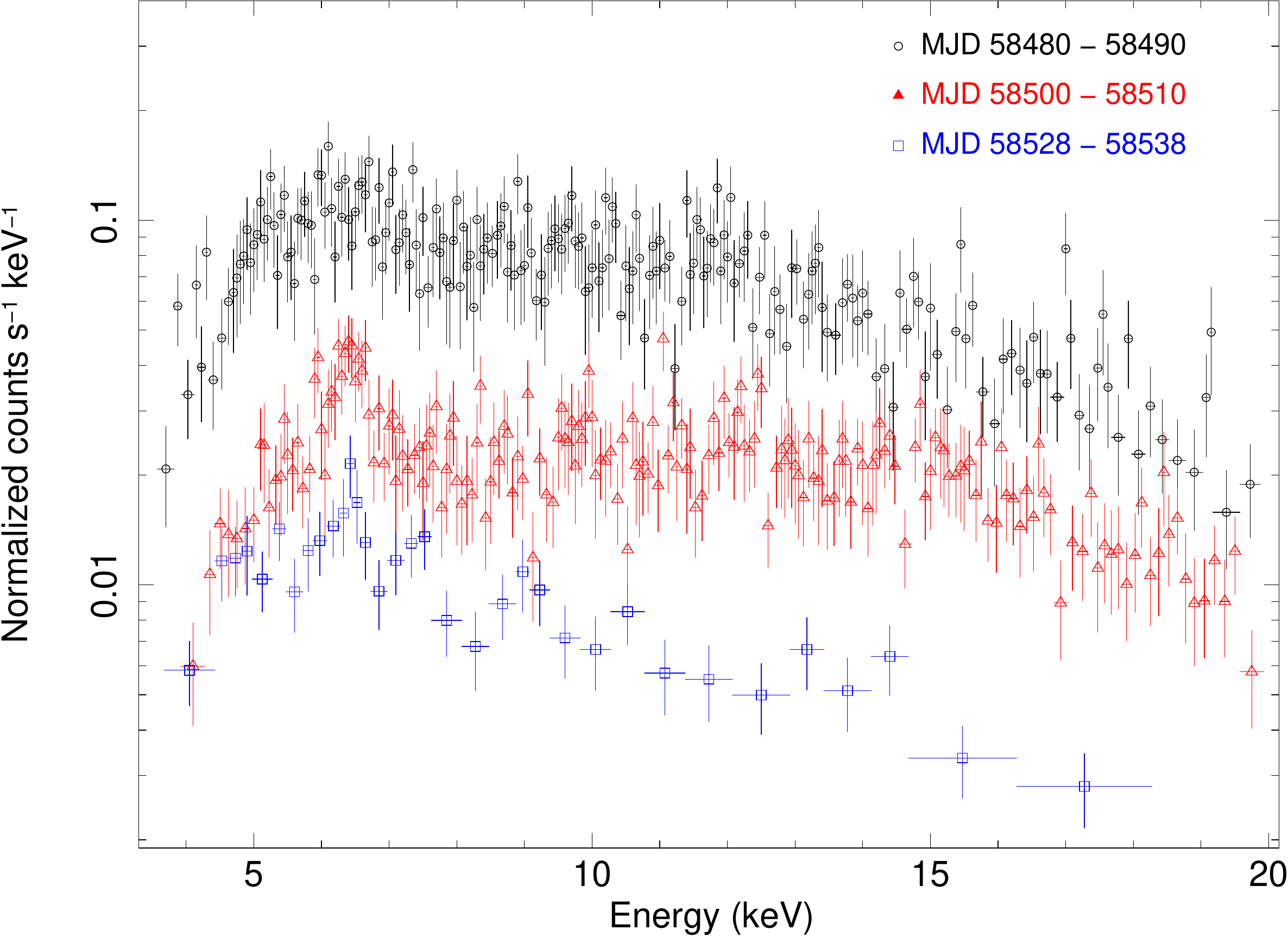}
    \caption{The figure shows the \textit{MAXI}/GSC spectra of GX 301-2 during three instances of the spin-up episode classified based on the strengths of N\textsubscript{H} and eqw\textsubscript{Fe}. The spectra during MJD 58528 to 58538 is shown in black ($\circ$) and scaled by a factor of 3 for better visibility. The spectra during MJD 58480 to 58490 is shown in red ($\triangle$) and MJD 58500 to 58510 in blue ($\square$), and both are shown in original scales.}
    \label{fig:spinup_spectra_3regions}
\end{figure}

As pointed out in the section \ref{intro}, the aim of this study is to identify any changes in the circum-binary/reprocessing environment of the pulsar during the spin-up episode. Using pointed observations, \citealt{nabizadeh2019} (\textit{NuSTAR}) and \citealt{liu2021disc} (\textit{Insight-HXMT}) reported contrasting results on the difference in the nature of reprocessing environment during and outside the 2019 spin-up episode. The latter reported relatively low N\textsubscript{H} during the spin-up episode (disk-fed state) when compared to outside the spin-up (wind-fed state), but the former reported no major differences in the spectral parameters. However, it has to be noted that each of the pointed observations only probes a tiny fraction of the orbital phase, and GX 301-2 is known to exhibit significant changes across the binary orbit \citep{nazma2014}. Therefore it is preferable to compare the reprocessing environment of the pulsar using observations that probe similar orbital phase of the binary. Moreover, our analysis shows that the reprocessing environment underwent significant variation during the spin-up episode itself (See Figure. \ref{fig:spinup_spectra_3regions} and \ref{fig:compare_orb_spinup}). Following are our observations about GX 301-2 during the spin-up episode:
\begin{itemize}
\item Usually the spin-up episodes are accompanied by an increase in the flux, indicating an increase in the accreted matter due to the formation of a transient accretion disk around the NS \citep{ghosh1979ApJ...234..296G}. The spin-up rate (the time derivative of spin frequency obtained by fitting straight line segments on the Figure. \ref{fig:spin-up_episode}) is relatively higher during the first orbit at about $6.5\times10^{-12}$ Hz s\textsuperscript{-1}, incidentally where the X-ray flux is also high. During the second orbit, where the X-ray flux has not increased as such and the spin-up rate is also lower at about $2.3\times10^{-12}$ Hz s\textsuperscript{-1} \citep[Also see Figure. 1 of][]{liu2020spin}.
\item The onset of spin-up happened at the orbital phase where N\textsubscript{H} would have been low, and iron equivalent width would have been high in an average orbit. But at the onset of spin-up, the eqw\textsubscript{Fe} is low, and N\textsubscript{H} is high.
\item In an average orbit, both N\textsubscript{H} and eqw\textsubscript{Fe} peaks at/near the pre-periastron flare. However, during the rapid phase of the spin-up episode, this happens $\sim5$ days prior to the phase of pre-periastron flare (observed at $\sim$ MJD 58502). This is accompanied by an increase in the X-ray flux which is clearly seen in the \textit{Swift}/BAT light-curve. At the phase of the pre-periastron flare, the values of N\textsubscript{H} and iron equivalent are lower than what is expected from an average orbit.
\item Soon after the unusual increase in absorption and luminosity observed at $\sim$ MJD 58502 during the rapid phase of the spin-up episode and the subsequent pre-periastron flare, the pulsar went into the slow spin-up state with an associated decrease in the luminosity.
\item After the apastron passage towards the end of the spin-up episode, at around 58530 MJD, the iron line equivalent width has shown an abrupt increase to values of $\sim1000$ eV, which is not observed during an average orbit.
\item During the entire spin-up episode the values of peak N\textsubscript{H} ($\sim100\times10^{22}$ atoms cm\textsuperscript{-2}) are relatively higher than that during the average orbits ($\sim 50\times10^{22}$ atoms cm\textsuperscript{-2}). However, the average equivalent width of the iron line during the spin-up episode is less than half of the long-term average value of the same. This is perhaps due to the fact that, during the spin-up episode, the source was bright throughout a binary orbit, while in the long-term averaged data, the spectrum is dominated by the emission during the pre-periastron flare.
\end{itemize}

\section*{Acknowledgements}

This research has made use of \textit{MAXI} data provided by RIKEN, JAXA and the \textit{MAXI} team.
We acknowledge the use of public data from the \textit{Swift} data archive and GBM Accreting Pulsars Program (GAPP). We thank the reviewer for the constructive comments that helped us improve the manuscript.

%%%%%%%%%%%%%%%%%%%%%%%%%%%%%%%%%%%%%%%%%%%%%%%%%%
\section*{Data Availability}
All the data used in this study are publicly available. The MAXI spectra can be downloaded from the web tool MAXI on-demand process, and the spin histories can be downloaded from GBM Accreting Pulsar Histories.



\bibliographystyle{mnras}
\bibliography{mnras_template} % if your bibtex file is called example.bib

%\begin{thebibliography}{99}
%\end{thebibliography}

%%%%%%%%%%%%%%%%%%%%%%%%%%%%%%%%%%%%%%%%%%%%%%%%%%

%%%%%%%%%%%%%%%%% APPENDICES %%%%%%%%%%%%%%%%%%%%%

\appendix
%%%%%%%%%%%%%%%%%%%%%%%%%%%%%%%%%%%%%%%%%%%%%%%%%%

% Don't change these lines
\bsp	% typesetting comment
\label{lastpage}
\end{document}